\documentclass{amsart}

\usepackage{tikz}
\usepackage{listings}

\newtheorem{theorem}{Theorem}[section]
\newtheorem{proposition}[theorem]{Proposition}

\newtheorem{example}[theorem]{Example}

\begin{document}

\title{Interleaved computation for persistent homology}
\author{Mikael Vejdemo-Johansson}

\begin{abstract}
  We describe an approach to bounded-memory computation of persistent homology and betti barcodes, in which a computational state is maintained with updates introducing new edges to the underlying neighbourhood graph and percolating the resulting changes into the simplex stream feeding the persistence algorithm.

  We further discuss the memory consumption and resulting speed and complexity behaviours of the resulting algorithm.
\end{abstract}

\maketitle

\section{Introduction}
\label{sec:introduction}

Persistent homology has turned out to be a powerful tool in data analysis as well as several other fields. When used in application areas, the computation of Betti barcodes is usually done in an exploratory fashion, with no good a priori intuition for what parameters to use for the computation.

Among the publicly available software: both \textsc{Dionysus}, \textsc{jPlex}, and \textsc{javaPlex} all produce a computational interface in which a particular run is computed in its entirety, and the internal state subsequently discarded. This is a problem for exploratory computations, as restarting a computation has a high cost, and little, if any, information can be salvaged should the computational parameters chosen exceed the capabilities of the platform.

Here, we propose a solution to these issues: explicitly saving the internal state of the computation, and an incremental computational paradigm that generates simplices of the Vietoris-Rips complex in the order they appear in the simplex stream, and computes persistent homology interleaved with the generation of new simplices.

There are bounds on how memory efficient an implementation of Vietoris-Rips and persistent homology can be: since the persistence algorithm pairs simplices, it will be difficult to store any less than each simplex once. However, we can discard the Vietoris-Rips simplex stream as it is consumed by the persistence algorithm, and grow not much beyond the data needed to store the initial stream. This is a significant difference from any existing implementation, in which the entire stream is generated, and subsequently the entire persistence computation structure is generated as well.

\section{Incremental Vietoris-Rips construction}
\label{sec:incr-viet-rips}


We propose here a construction of the Vietoris-Rips simplex stream by an incremental process: assuming we can acquire a stream of edges in the Vietoris-Rips neighbourhood graph, we consume one edge at a time, and generate all simplices implied by any given edge just before they are consumed by the persistence algorithm. 


At the core of this approach is the observation that we can maintain a current list of maximal cliques, and any new maximal clique after adding an edge comes from the intersection of maximal cliques containing either of the two endpoints of the edge. Any maximal clique equal to a such intersection joined with one of the endpoints will vanish in the update, and any other maximal cliques stays --- as do any maximal among the cliques constructed as such adjoined intersections.

\begin{proposition}
  Suppose $\mathcal L$ is a system of maximal cliques in a graph $\Gamma$. Create the graph $\Gamma\cup e$ adjoined with the new edge $e=(s,t)$.  Write $\mathcal L_v$ for the maximal cliques in $\Gamma$ containing the vertex $v$.

  If $\ell_s\in\mathcal L_s$ and $\ell_t\in\mathcal L_t$, then $(\ell_s\cap\ell_t)\cup\{s,t\}$ is a clique in $\Gamma\cup e$.

  Maximal cliques in $\Gamma\cup e$ are cliques in $\mathcal L$ or on the shape $(\ell_s\cap\ell_t)\cup\{s,t\}$ that are not contained in any other cliques in this list.
\end{proposition}
\begin{proof}
  There are a few statements to be proven here.
  \begin{enumerate}
  \item $(\ell_s\cap\ell_t)\cup\{s,t\}$ is a clique. Indeed, any vertex in $\ell_s\cap\ell_t$ is djacent to both $s$ and $t$. Furthermore, these vertices are all adjacent to each other. Hence, $(\ell_s\cap\ell_t)\cup\{s,t\}$ is a clique in $\Gamma\cup e$.
  \item Any maximal clique in $\Gamma\cup e$ is either maximal in $\Gamma$ or has the shape $(\ell_s\cap\ell_t)\cup\{s,t\}$. Indeed, suppose $\ell$ is a maximal clique in $\Gamma\cup e$. 

If $e\not\in\ell$, then $\ell$ is a clique in $\Gamma$. Suppose $\ell$ is not maximal in $\Gamma$. Then there is some larger clique $\ell'$ containing $\ell$. However, $\ell'$ is still a clique in $\Gamma\cup e$, which contradicts the maximality of $\ell$ in $\Gamma\cup e$. 

Suppose now that $e\in\ell$. Then $\ell\setminus\{s,t\}$ is a clique in $\Gamma$. We claim that $\ell\setminus s$ (and $\ell\setminus t$: the argument is symmetric) are contained in maximal cliques in $\Gamma$ such that the intersection of these maximal cliques is exactly $\ell\setminus\{s,t\}$. It is immediately clear that $\ell\setminus s$ is contained in at least one maximal clique. Indeed, since $\ell\setminus s$ avoids $e$, all its edges are present in $\Gamma$, and thus $\ell\setminus s$ is a clique in $\Gamma$. Similarily, $\ell\setminus t$ is contained in at least one maximal clique. Remains to show that $\ell\setminus\{s,t\}$ is the intersection of some pair of such maximal cliques. Suppose there was no such pair that had this as their precise intersection. Then, all intersections between maximal cliques containing either are larger than $\ell$. Pick some pair of maximal cliques $\ell_s\supseteq\ell\setminus t$, $\ell_t\supseteq\ell\setminus s$. Then $(\ell_s\cap\ell_t)\cup\{s,t\}$ is a clique in $\Gamma\cup e$ by the same reasoning as above. Furthermore, $(\ell_s\cap\ell_t)\cup\{s,t\}$ strictly contains $\ell$. This contradicts the maximality of $\ell$ and thus finishes the proof.
  \end{enumerate}
\end{proof}

Suppose now we have a set of maximal cliques $\mathcal L$ for $\Gamma$ and a set of maximal cliques $\mathcal L'$ for $\Gamma\cup e$.  Then the set of simplices introduced by the introduction of $e$ corresponds to the set of cliques in $\Gamma\cup e$ that were not cliques in $\Gamma$. Obviously, any cliques in $\mathcal L'$ containing $e$ will be such new simplices, but so will also any subcliques of these cliques that contain $e$.

Indeed, any clique containing $e$ was obviously not present in $\Gamma$. Conversely, any clique introduced by the addition of $e$ has to contain $e$.

\begin{example}
  Consider the graph: \\
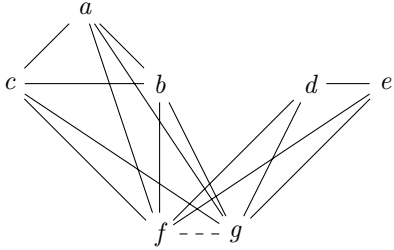
  \begin{tikzpicture}
    \node (a) at (0,3) {a};
    \node (b) at (1,2) {b};
    \node (c) at (-1,2) {c};
    \node (d) at (3,2) {d};
    \node (e) at (4,2) {e};
    \node (f) at (1,0) {f};
    \node (g) at (2,0) {g};
    \draw (a) -- (b) -- (c) -- (a)
    (d) -- (e)
    (g) -- (a) -- (f) -- (b) -- (g) -- (c) -- (f) -- (d) -- (g) -- (e) -- (f);
    \draw [dashed] (g) -- (f);
  \end{tikzpicture}

  This graph has maximal cliques $abcf$, $abcg$, $def$ and $deg$. Adding the edge $f-g$ yields the new cliques $abcfg$ from the intersection $abcf \cap abcg$, $fg$ from the intersections $abcf\cap deg$ and $def\cap abcg$ and the clique $defg$ from the intersection $def\cap deg$. We may recognize $fg$ as redundant, since it is contained in (several of) the other candidate cliques. All previous maximal cliques are contained in one of the remaining maximal cliques $abcfg$ and $defg$.

Furthermore, we may consider faces of $abcfg$ and $defg$. As soon as either $f$ or $g$ are removed from a clique, we find a clique of $\Gamma$. Thus, the new faces are the boolean algebras formed by pointwise union of the powerset of $\{a,b,c\}$ or $\{d,e\}$ respectively with the set $\{f,g\}$. Thus, in this case, the new faces are $fg, afg, bfg, cfg, dfg, efg, abfg, acfg, bcfg, defg, abcfg$.
\end{example}

We note that this implies an efficient algorithm for generating new simplices: for each clique created by adjoining the new edge to a clique intersection, we generate all subsets of the intersection clique, and add the edge to each thus generated subset. Finally, we remove duplicates from the union of all such simplices.

\begin{lstlisting}
  max_cliques = [[v] for v in vertices]
  def processEdge(source, target):
    s_cliques = [cl for cl in max_cliques if source in cl]
    t_cliques = [cl for cl in max_cliques if target in cl]
    intersections = [ intersection(cls,clt) 
                        for cls in s_cliques 
                        for clt in t_cliques ]
    max_intersections = [ cl in intersections
                             if cl is maximal in intersections ]
    new_cliques = [ union(cl, [source, target])
                      for cl in max_intersections]
    old_cliques = [ cl for cl in max_cliques 
                       if 1 < min [ size(difference(cl,ncl))
                                      for ncl in new_cliques ] ]
    simplex_seeds = unique([ subsets(cl)
                               for cl in max_intersections ])
    new_simplices = [ union(cl, [source, target])
                        for cl in simplex_seeds ]
    return unique(new_simplices)
\end{lstlisting}

\section{Interleaved algorithm}
\label{sec:interl-algor}

The persistence algorithm, as described in \cite{elz2000} and subsequently refined in \cite{cz2005}, consumes a stream of simplices, and maintains a state from which at any time both the finished intervals as well as the half-open intervals still unfinished can be read. We propose to run the persistence algorithm interleaved with the incremental Vietoris-Rips algorithm above, thus discarding any information not needed for the subsequent computation, and staying continuously with a state that can be easily interpreted at any point along the computation as well as used for continued computation.

\section{Space usage}
\label{sec:space-usage}

The current best practice for Vietoris-Rips complex construction was described in \cite{z2010vr}. In this approach, first a filtered graph is generated. Then, from this graph, we generate a Vietoris-Rips complex, and then from this complex, a persistence barcode with the table in the persistence algorithm.

A full distance graph will take $O(n^2)$ space: at least three values (source index, target index and distance) needs to be stored for each of the $n\choose 2$ edges available on $n$ vertices.

A full Vietoris-Rips complex, in turn, will consume exponential space in the number of vertices. However, most analyses work with a $k$-skeleton for some small $k$, and indeed, this research was originally motivated by attempts to compute degree 2 homology, and thus working with the 3-skeleton instead of the more widespread 2-skeleton. This has a memory consumption of $O(n^k)$, since $\sum_{r=1}^k {n\choose r}$ tuples need to be stored, and each such $r$-tuple requires $r$ indices.

The homology computation, finally, erects a table with \emph{marked simplices} and their \emph{cascades}, as described in \cite{cz2005}. This table stores each simplex that creates a homology class, and for each such simplex, some number of cascade simplices. It is not known, currently, what the behaviour of the cascade is in the algorithm: what the expected distribution of cascade storage for a generic barcode computation would be. However, since the representatives of any homology class can be found in the cascades of paired marked simplices, the cascade storage sizes correspond to the storage sizes of generic representatives for homology classes.

Finally, the barcode produced by the persistence algorithm stores at least a start and end time for each interval, creating a storage consumption of $O(n^k)$.

All in all, we are faced with a memory consumption of $O(n^k)$ for the entire computation.

There are some ways we can try to accomodate the storage needs of a computation. One of them is to use laziness in our computations, and only generate the things we are about to consume. The $O(n^k)$ storage needs for the Vietoris-Rips complex can be broken up into pieces that are far smaller, in fact of a space complexity dependent on the number of new cofaces of any given edge rather than on the entire simplicial complex. We still face $O(n^k)$ in storage needs for the persistence algorithm table, but the constant is significantly smaller in this latter case.

Another way to approach this problem is to use the stratified nature of modern computers. We have many different kinds of memory: L1 and L2 caches, RAM memory, swapped out RAM pages, explicit file storage, networked storage. By writing out anything not explicitly needed for future computation to disk, we can relieve the load on RAM (currently ranging in sizes from 2G to 64G in most hardware available to individual researchers) and instead put this load on hard drives (currently ranging in sizes from 250G to multiple tera-byte in corresponding hardware).

Candidates for relegation to slower storage in the pipeline we describe here are:
\begin{enumerate}
\item The distance graph. We can generate all $O(n^2)$ pairwise distances, and write them to disk as we find them. These can then be sorted, if necessary in-place on disk, to yield a file that stores, say, three values for each of the $n\choose 2$ distances in question, namely distance, index to the source vertex and index to the target vertex.

Once the sorted distance graph is stored, it is a simple disk read of a finite and small amount of data to gain access to the next edge in the Vietoris-Rips graph.
\item The persistence intervals. Once a persistence interval is closed, the information about the actual interval is not needed for the further computation. The corresponding marked simplex and cascade are needed, but the tuple of birth-time, death-time and optionally the associated homology cycle are not needed for further work.
\end{enumerate}

We can not escape the $O(n^k)$ storage requirements. However, by minimizing memory residence for any of the irrelevant steps -- either by using an incremental algorithm, or by disk residence -- will stretch the reach of current persistent homology work.

These ideas: on-disk storage of the generating graphs and of the resulting intervals, as well as minimal memory residence for the Vietoris-Rips complex before inclusion in the persistence tables, are in no way isolated to this particular implementation; we expect that all existing implementations will benefit from these ideas.

\bibliographystyle{apalike}
\bibliography{../library.bib}

\end{document}